
\input phyzzx
\overfullrule=0pt
\def\kkbar{$\ko$--$\kob$}
\def\ko{K_0}
\def\kob{{\bar K}_0}
\def\refitem#1{\r@fitem{#1)}}
\def\REFWRITE{\R@FWRITE\rel@x}

\def\lash#1{#1\!\!\! / }

\def\ml{m_L}
\def\ms{m_S}
\def\gl{\Gamma_L}
\def\gs{\Gamma_S}

\def\ek{\epsilon}
\def\dk{\Delta}
\def\d{d}
\def\gam{\Gamma}
\def\dm{\Delta m}
\def\dg{\Delta \gam}
\def\kl{K_L}
\def\ks{K_S}
\def\rk{\rho_K}

\def\rl{\rho_L}
\def\rs{\rho_S}
\def\ri{\rho_I}
\def\rib{\rho_{\bar I}}
\def\gl{\gam_L}
\def\gs{\gam_S}
\def\\ml{m_L}
\def\ms{m_S}
\def\gb{{\bar \gam}}

\def\egl{e^{-\gl\tau}}
\def\egs{e^{-\gs\tau}}
\def\egi{e^{-\gb\tau}}

\def\em{e^{-i\dm\tau}}
\def\emb{e^{+i\dm\tau}}

\def\lp{\pi^{-}\ell^{+}\nu}
\def\lm{\pi^{+}\ell^{-}\nu}
\def\bed{{\beta \over \d}}
\def\beds{{\beta \over \d^{\star}}}

\def\etapp{\eta_{+-}}

\def\phipp{\phi_{+-}}

\def\gdg{{\gamma \over \dg}}

\def\P{{\cal P}}
\def\PP{\bar{\cal P}}

\def\Dt{\Delta\tau}
\def\O{{\cal O}}

\def\ra{\rightarrow}
\def\Re{{\rm Re}}
\def\Im{{\rm Im}}
\def\qmg{{(\diamondsuit)}}

\def\cpt{{\it CPT\/}}

\def\cp{{\it CP\/}}
\def\cpv{\cp\ violation}

\def\phif{$\phi\,$ factory}
\def\phifs{$\phi\,$ factories}
\def\qm{quantum mechanics}

\REF\us{ P.Huet and M.E. Peskin, preprint SLAC-PUB-6454 (1994).  }
\REF\pdb{Particle Data Group, K. Hikasa et al.,
Review of Particle Properties, Phys. Rev. {\bf D45} (1992). }
\REF\hawk{S.W. Hawking, Phys. Rev. {\bf D 14}, 2460 (1975); Commun.
math. Phys. {\bf 87}, 395 (1982).}
\REF\dpage{D.N. Page, \sl Gen. Rel. Grav. {\bf 14,} \rm (1982);
L. Alvarez-Gaum\'e and C. Gomez, \sl Commun. Math. Phys. {\bf 89,}
 \rm 235 (1983);
 R. M. Wald,{\it General Relativity}, Chicago, Ill., Univ. of Chicago
Press, 1984.}
\REF\ellishns{J. Ellis, J.S. Hagelin, D.V. Nanopoulos and M. Srednicki,
Nucl. Phys. {\bf B241},381 (1984).}
\REF\ellis{J. Ellis , N. E. Mavromatos and D. V. Nanopoulos, \sl
Phys.Lett.
{\bf B293}, \rm 142 (1992).}
\REF\ellistwo{J. Ellis, N. E. Mavromatos, and D. V. Nanopoulos,
 CERN-TH.6755/92 (1992).}
\REF\bps{T. Banks, M. Peskin and L. Susskind, Nucl. Phys. {\bf B 244}
125 (1984).}
\REF\pecceione{C. D. Buchanan, R. Cousin, C. O. Dib, R. D. Peccei and J.
Quackenbush, \sl Phys. Rev. {\bf D45}, \rm 4088 (1992).}
\REF\peccei{For a review, see R.D. Peccei, preprint
UCLA/93/TEP/19 (1993).}
\REF\pecceiall{C.O. Dib and R.D. Peccei, Phys.Rev. {\bf D 46},2265
(1992).}
\REF\frascati{{\it The DA$\Phi$NE PHYSICS HANDBOOK},  edited by L.
Maiani, G. Pancheri and N. Paver, (INFN, Frascati). }
\REF\heidi{C. Geweniger, \etal.,  Phys. Lett. {\bf 48 B}, 483  (1974).}
\REF\heidii{C. Geweniger, \etal,
 \sl Phys. Lett. {\bf 48 B}, 487 (1974).}
\REF\heidiii{S. Gjesdal, \etal,  \sl Phys. Lett. {\bf 52 B}, \rm
113 (1974).}
\REF\cplear{CPLEAR collaboration, R. Adler et al., \sl Phys. Lett. {\bf B
286}, \rm 180 (1992).}
\REF\maianirev{L. Maiani, in the DA$\Phi$NE Handbook, ref. \frascati,
         vol. I.}
\REF\dunietz{I. Dunietz, J. Hauser and J. Rosner, Phys. Rev. {\bf D
35}, 2166 (1987).}

\pubnum={6491}
\date={May 9, 1994}
\pubtype={T/E}
\titlepage
\title{Testing Violation of CPT and Quantum Mechanics in
the \kkbar\ System}
\author{Patrick Huet}
\SLAC
\abstract{I report a recent study \refmark\us made in collaboration
with M.E. Peskin, on the time dependence of a kaon beam propagating
according to a generalization of quantum mechanics due to Ellis,
Hagelin, Nanopoulos and Srednicki, in which \cp- and \cpt-violating
signatures arise from the evolution of pure states to mixed states.
Constraints on the magnitude of its parameters are established on the
basis of existing experimental data. New facilities such as \phifs\
are shown to be particularly adequate to study this generalization
from \qm\ and to disentangle its parameters from other \cpt\
violating perturbations of the kaon system.}

\centerline {\it To appear in Proceedings}
\centerline {\it of the First International Conference}
\centerline {\it on Phenomenology of Unification from}
\centerline {\it Present to Future, Roma, ITALY -- March 23-26,1994}
\endpage
 \chapter{Introduction}

Developments in the quantum theory of gravity have led S. Hawking to
propose a generalization of quantum mechanics which allows the
evolution of pure states to mixed states.\refmark\hawk This
formulation was shown by D. Page to be in conflict with \cpt\
conservation.\refmark\dpage
Ellis, Hagelin, Nanopoulos, and Srednicki (EHNS)\refmark\ellishns
were first to observe that systems which exhibit quantum coherence
over a macroscopic distance are most appropriate to probe the
violation of quantum mechanics of the type proposed by S. Hawking.
One of the simplest system exhibiting this property is a beam of
neutral kaons. EHNS set up a generalized evolution equation for the
\kkbar\ system in the space of density matrices which contains three
new \cpt\ violating parameters $\alpha$,$\beta$ and $\gamma$. These
parameters have dimension of mass and could be as large as
$m_K^2/m_{\rm Pl} \sim 10^{-19}$ GeV.

  Recently, Ellis, Mavromatos, and Nanopoulos
(EMN)\refmark{\ellis,\ellistwo} reconsidered this analysis.
Exploiting experimental data on $\kl$ and $\ks$, they presented an
allowed region in the space of the parameters $\alpha$, $\beta$, and
$\gamma$. This region is compatible with the expected order of
magnitude above and with the possibility that violation of quantum
mechanics accounts for all \cp\ violation observed in the \kkbar\
system. However, as these authors were aware, this analysis does
not include data from the intermediate time region of the kaon beam.
This time region has proven to be the main source of constraints on
quantum mechanical \cpt\ violating perturbations.\refmark\peccei

 The present talk is based on a recent work\refmark\us done in
collaboration with M.E. Peskin, which develops a general
parameterization to incorporate \cpt\ violation from both within and
outside quantum mechanics and uses it to analyze past, present and
future experiments on the \kkbar\ system. Studies of the time
dependence of the kaon system of the early 1970's are combined with
recent results from CPLEAR to constrain the EHNS parameters $\beta$
and $\gamma$, limiting the contribution of violation of \qm\ to no
more than $10 \%$ of the \cp\ violation observed in the \kkbar\
system.

 \phifs\ are new facilities\refmark\frascati dedicated to the
study of the properties of the kaon system; they are expected to give
particularly incisive tests of \cpt\ violation.\refmark\peccei
It is shown that these future facilities are especially suitable
to test violation of \qm\ and to disentangle the EHNS parameters from
other \cpt\ violating perturbations.

\chapter{general formalism}
\section{\undertext{Quantum-mechanical evolution}}
In this section we describe the quantum mechanical time evolution of
a beam of kaon and generalize it to allow a pure state to
evolve to a mixed state.

 Let $\rk(\tau)$ be the density matrix of a kaon beam at proper time
$\tau$. Its value at the source is fixed by the production mechanism.
Its values along the beam are then obtained from
an effective hamiltonian $ H\,\, = M - {i\over 2} \Gamma,$ which
incorporates the natural width of the system, and the laws of quantum
mechanics
 $$   i {d\over d\tau} \rk =    H\,\rk - \rk\, H^{\dagger} .
\eqn\densitymateq
$$
The solution of \densitymateq\ is generally expressed in terms of the
properties of $\kl$ and $\ks$ as
 $$\rk(\tau)\,=\,A_L \rl^\qmg
\egl+A_S\rs^\qmg \egs+A_I\ri^\qmg \egi\em+
A_{\bar I}\rib^\qmg \egi\emb.
\eqn\rhotau$$
The coefficients $A_{L,S}$ and $A_{I,{\bar I}}$ describe the
initial conditions of the beam and we defined
 $$ \gb= {\gl + \gs \over 2}\,\, ,\qquad
\dg= \gs - \gl\qquad {\rm and} \qquad \dm = \ml-\ms\, .
\eqn\dmandgdef
 $$
For later purpose, it is convenient to introduce the complex quantity
 $$ \eqalign{   d &= \dm + {i\over 2} \dg = \bigl((3.522\pm 0.016) + i
                   (3.682 \pm 0.008) \bigr) \times 10^{-15} {\rm GeV}
                   \cr
               d &= |d| e^{i(\pi/2-\phi_{SW})} \qquad
         \phi_{SW}=(43.73\pm 0.15)^\circ \cr} \, .
\eqn\superwdef$$

 The density matrices $\rl^\qmg$, $\rs^\qmg$ and $\ri^\qmg$,
$\rib^\qmg$ are expressible in terms of the pure states $\ket{\kl}$
and $\ket{\ks}$ as follows
 $$
\eqalign{
\rl^\qmg\,=&\,\ket{\kl}\bra{\kl}  \cr
\rs^\qmg\,=&\,\ket{\ks}\bra{\ks}   \cr}\qquad\qquad
\eqalign{
\ri^\qmg\,=&\, \ket{\ks}\bra{\kl}   \cr
\rib^\qmg\,=&\,\ket{\kl}\bra{\ks}.  \cr}
\eqn\eigenvectors
$$
The state $\ket{K_{L(S)}}$ is the sum of the \cp\ {\it even(odd)}
state $\ket{K_{1(2)}}$ and a small \cp\ {\it odd(even)} component
proportional to the state $\ket{K_{2(1)}}$
 $$  \eqalign{
 \ket{K_S} & = N_S \bigl( \ket{K_1}
                    + \ek_S\ket{K_2} \bigr) \crr
 \ket{K_L} & = N_L \bigl( \ket{K_2}
                    + \ek_L\ket{K_1} \bigr) \cr}
\eqn\shortlongdef$$
where
$ \ek_S = \ek_M + \dk $ and $ \ek_L = \ek_M - \dk$.
 The parameter\foot{We use the notation of ref. \maianirev .} $\ek_M$
is odd under \cp\ but even under \cpt\ while the parameter  $\dk$ is
odd under both \cp\ and $\cpt$. $N_S$, $N_L$ are real, positive
normalization factors.

Any observable of the kaon beam can be computed by tracing $\rk$ with
an appropriate operator  ${\cal P}$, we write
 $$  \VEV{{\cal P}} = \tr \bigl[ \rk  \O_\P \bigr].
\eqn\generalVEV
 $$
This expression for expectation values will remain true in the
generalization of quantum mechanics described below. The most
relevant observables are the semi\break leptonic decays $K
\rightarrow \pi^\pm\ell^\mp\nu$ and the charged and neutral pion
decay $K \rightarrow \pi^+\pi^-,\,\, \pi^0\pi^0$. The operators for
the semileptonic decays are
 $$  \O_{\ell^\pm}   =  |a|^2 \ket{K_0} \bra{K_0} = {|a \pm b|^2\over 2}
                            \pmatrix{ 1 &  \pm1 \cr \pm1 & 1\cr},
\eqn\Operatordecay
 $$
and for the 2 pion decays, they are
 $$  \O_{+-}   =  |X_{+-}|^2
          \pmatrix{ 1 &  Y_{+-} \cr Y^*_{+-} & |Y_{+-}|^2\cr}\ ,\qquad
 \O_{00}   =  |X_{00}|^2
   \pmatrix{ 1 &  Y_{00} \cr Y_{00}^* & |Y_{00}|^2\cr},
\eqn\Opptwo
$$
where
$$   X = \VEV{\pi\pi \Bigm| K_1 } \ ,\qquad
Y =   { \VEV{\pi\pi \bigm| K_2} \over\VEV{\pi\pi \bigm| K_1}} .
\eqn\Oppdefs
$$
More explicitly,\refmark\us\
$$ \eqalign{
Y_{+-} &= \bigl( {\Re B_0\over  A_0}
                        \bigr) + \ek' \cr
Y_{00} &= \bigl( {\Re B_0\over  A_0}
                        \bigr) -2 \ek' \cr}\, .
\eqn\Ydefs
$$
The quantities $\Re(B_0/A_0)$ and $\Re(b/a)$  parameterize
\cpt-violating decay amplitudes\refmark\peccei while $\ek'$ is the
standard \cp\ violating parameter in the pion decay. To illustrate
how to use these operators, we compute the charged pion decay rate at
large time $\tau \gg 1/\gs$ in the evolution of the beam
 $$\eqalign{ \Gamma(\kl \rightarrow \pi^+\pi^-)
 &\propto\, \tr  \rl \O_{+-} \cr
 &\propto\, |\ek_L|^2 +|Y_{+-}|^2+ 2 \Re\, Y_{+-}\ek_L^* \cr
 &\propto\, |\etapp^\qmg|^2 \cr} \, ,
 \eqn\examppp
 $$
where $\etapp^\qmg$ is the complex number $\ek_L + Y_{+-} =
\ek_L+\Re(B_0/A_0)+\ek'$.

\section{\undertext{Generalized time evolution}}
Violation of quantum mechanics is clearly a small perturbation of the
formalism described above. EHNS proposed to account for the loss of
coherence in the evolution of the beam by adding terms on the RHS of
\densitymateq\ which preserves the linearity of the time-evolution
 $$   i {d\over d\tau} \rk =    H\,\rk - \rk\, H^{\dagger}
\,+\,\lash{\delta h} \,\rk .
\eqn\densitymateqnew
$$
Requirements that such terms do not break conservation of probability
and do not decrease the entropy of the system make $\lash{\delta h}$
expressible in terms of six parameters. In order to lower this number
to a more tractable one, EHNS further assume that these terms
conserve strangeness, reducing $\lash{\delta h}$ to three unknown
parameters $\alpha$, $\beta$ and $\gamma$.

The solution of \densitymateqnew\ has the general form given in
\rhotau\ but with the eigenmodes $\rl$, $\rs$ and $\ri$ changed
to,\foot{We use a diamond superscript to label quantum mechanical
quantities.} in first order in small quantities
 $$
\rl= \rl^\qmg +  \gdg  \rs^\qmg + \bed \ri^\qmg + \beds \rib^\qmg
\eqn\rlwritten
$$
$$
\rs= \rs^\qmg -  \gdg  \rl^\qmg - \beds \ri^\qmg - \bed \rib^\qmg
\eqn\rswritten
$$
$$
\ri= \ri^\qmg  -\beds \rs^\qmg + \bed \rl^\qmg -{i\alpha \over 2\dm}
\rib^\qmg
\eqn\riwritten
$$
$$
\rib= \ri^\qmg  -\bed \rs^\qmg + \beds \rl^\qmg +{i\alpha \over 2\dm}
\ri^\qmg \, .
\eqn\ribwritten
$$
The corresponding eigenvalues are corrected  by the shifts
$$\matrix{
   \gl \ra \gl + \gamma \ , & \gs \ra \gs + \gamma\ , \cr
       \gb \ra \gb + \alpha \ , &
 \dm\ra         \dm\cdot (1-{1\over 2}(\beta/\dg)^2)\ .  \cr }
\eqn\eigenvshifts$$
The shifts of $\Gamma_L$, $\Gamma_S$, and $\dm$ can be absorbed by
redefinition of these parameters.  The shift of $\dm$ is of relative
size $10^{-6}$ and so is negligible in any event.  If we redefine
$\bar \Gamma$ to be the average of the new $\Gamma_S$ and $\Gamma_L$,
then the interference terms $\ri$ and $\rib$ fall off at the rate $
\bar\Gamma  + (\alpha - \gamma)$. This correction is not relevant to
current experiments unless $\alpha$ is as large as $10^{-2}\gb$; in
that case $\alpha$ would be 10 times larger than  the familiar
\cp-violating parameters.

The major effect of violation of quantum mechanics is embodied in the
eigenmodes $\rl$, $\rs$, $\ri$. These density matrices are no longer
pure density matrices in contrast to their quantum mechanical
counterparts. This loss of purity alters the decay properties of the
beam. For example, the properties of the beam at large time, ${\tau \gg
1/\gs}$, are dominated by the properties of $\rl$.  The second term on
the RHS of \rlwritten\ is proportional to $\rs^\qmg$ and is even under
\cp\ conjugation. As a result, we expect an enhancement in the decay
into 2 pion at late time in the evolution of the beam. More
specifically, in leading order in small quantities, we find for the
decay rate at large time
 $$\eqalign{ \Gamma(\kl\rightarrow \pi\pi)
 &\propto\, \tr  \rl \O_{\pi\pi} \cr
 &\propto\, \tr \rl^\qmg \O_{\pi\pi} + \gdg \tr \rs^\qmg \O_{\pi\pi}
 + 2 \Re \, \Bigl( \bed \tr \ri^\qmg \O_{\pi\pi}\Bigr) \cr
 &\propto\, |\etapp^\qmg|^2 + \gdg + {\rm higher}\,{\rm order} = R_L  \cr} \,
,
\eqn\decayrate
 $$
There is, indeed, an enhancement of the 2 pion decay rate
proportional to $\gdg$. Similarly, in the intermediate time region ($
\tau \sim 1/\gb$), the dominant contribution to the 2 pion decay
comes from $\ri$ and its hermitian conjugate $\rib$. From its
expression \riwritten , $\ri$ has a piece proportional to $\rs^\qmg$
which shifts the 2 pion decay rate by an amount $\propto
\beta/|d|\,\cos\phi_{SW}$ and $\beta/|d|\,\sin\phi_{SW}$ in the
intermediate time region.

{}From the simple arguments above, we expect the phenomenology of the
kaon beam to be affected in leading order in the parameters $\bed$
and $\gdg$. In particular, we expect corrections of order $\bed$ in
the intermediate time region. We will exploit this fact in the next
section to establish some experimental bounds on violation of quantum
mechanics.

\chapter{Experimental constraints on $\alpha$ and $\beta$}

This section is a summary of an analysis made in ref. \us\ which
establishes constraints on the EHNS parameters using present
experimental data. In this analysis we combine very accurate
measurements on the time dependent kaon system
made in the early 1970's by the CERN-Heidelberg
collaboration\refmark\heidi with new data from the CPLEAR
experiment.\refmark\cplear
Provided that the EHNS parameters do not accidentally cancel against
the effects of the \qm\ \cpt\
violating parameters in direct decay $Y$ and $ b/a$
introduced in \Operatordecay-\Ydefs , we are able to give
stringent bounds on $\beta$ and $\gamma$ which limit their effects to
be at most $10\%$ of the observed \cp\ violation in the \kkbar\
system.
To simplify the following discussion, we temporarily neglect \qm\
\cpt\ violation in direct decays: $Y=b/a=0$. We will reintroduce them
at the end of this section.

For this analysis, we need two observables. the time dependent 2 pion
decay rate
 $$ \eqalign{ {\Gamma(K(\tau)\rightarrow \pi\pi)
 \over   \Gamma(K(0)\rightarrow \pi\pi)}=&
  {\tr \rk(\tau)\O_{+-}\over  \tr \rk(0)\O_{+-} } \cr
=& e^{-\gs\tau}+2|\etapp|^2\cos(\dm\tau-\phipp) + R_L e^{-\gl\tau} \cr}
\eqn\obspp
$$
and the semileptonic decay rate at large time $\tau\gg 1/\gs$
 $$\eqalign{
 \delta_L =& {\Gamma(\kl\rightarrow \lp)-\Gamma(\kl\rightarrow \lm) \over
   \Gamma(\kl\rightarrow \lp)+\Gamma(\kl\rightarrow \lm)} \cr
  =& {\tr \rk(\tau) \bigl(\O_{\ell^+} - \O_{\ell^-}\bigr) \over
\tr \rk(\tau) \bigl(\O_{\ell^+} + \O_{\ell^-}\bigr) }
 \cr} \, .
\eqn\obsrhodef
$$
The relevant measurable quantities are $R_L$, $\delta_L$ and
$\etapp$. $R_L$ and $\delta_L$ reflect the properties of the beam
which evolved at large time; the complex number
$\etapp=|\etapp|\exp(i\phipp)$ is a property of the intermediate time
region.

In \qm, they are related according to
$$\eqalign{
 R_L &= |\etapp|^2 \cr
 {\delta_L \over 2} &= \Re\, \etapp \cr} \, .
 \eqn\qmrelation
$$
After allowance has been made for \qm\ violation, they relate
according to
 $$\eqalign{
 R_L &\simeq |\etapp|^2 + \gdg + 4 {\beta \over |d|}|\etapp|\cr
 {\delta_L \over 2} &= \Re\, (\etapp - 2\bed)\cr} \, .
 \eqn\newrelation
 $$
The corrections of order $\beta$, $\gamma$ are the ones we described
in the previous section but including some second order corrections.
The geometry of these corrections is given in Fig. 1a . The current
experimental situation is discussed in ref. \us\ and shown on Fig. 1b.

According to equations \newrelation, the parameter $\beta$ is
proportional to the distance of the ellipse to the vertical band
while the distance of the ellipse to the arc provides a measurement of
$\gamma$. This comparison leads to the bounds
 $$\eqalign{
    \beta =&  (0.12 \pm 0.44) \times 10^{-18}\ {\rm GeV} \cr
   \gamma =&  (- 1.1 \pm 3.6) \times 10^{-21} \ {\rm GeV} \cr}.
\eqn\valofgamma
$$
To obtain these bounds, we set to zero the \cpt\ quantum mechanics
perturbations of the decay amplitude ${b/ a}$ and $B_0/ A_0$
introduced in \Operatordecay-\Ydefs. If we restore these parameters
in the above analysis, we find instead
 $$ \eqalign{   \beta&  + {|d|\over \sin\phi_{SW}}\Re\bigl(
         {b\over a} - {B_0\over A_0} \bigr)
= (0.12 \pm 0.44) \times 10^{-18}\ {\rm GeV} \cr
   &\gamma -  {|d|\over \sin\phi_{SW}}\Re\bigl(
         {b\over a} - {B_0\over A_0} \bigr)   =
         (- 1.1 \pm 3.6) \times 10^{-21} \ {\rm GeV}  \cr}\,  .
\eqn\betagammatwo$$
 Thus, our previous constraints on $\beta$ and $\gamma$ now appear as
constraints on combinations of \cpt-violating parameters. Unless,
unnatural cancellations occur among these \cpt\ violating parameters,
they can be independently constrain, in which case, neither of them
contributes more than $10\%$ of the total \cp\ violation
observed in the \kkbar\ system.

With we advent of a new generation of facilities such as \phifs,
which can perform incisive tests on \cpt\ violation,\refmark\frascati
come new ways of probing \qm. This is the object of the next section.

\chapter{Tests of quantum mechanics at a $\phi$-factory}

At a \phif, a spin-1 meson decays to an antisymmetric state of two
kaons which propagates with opposite momentum.  If the kaons are
neutral, we can describe the resulting wavefunction, in the basis of
\cp\ eigenstates $\ket{K_1}$, $\ket{K_2}$, as
 $$
\phi\rightarrow{1\over {\sqrt 2}}
\Bigl( |K_1,p>\otimes|K_2,-p>-  |K_2,p>\otimes|K_1,-p> \Bigr)\  .
\eqn\phistate
 $$

The two-kaon density matrix resulting from this decay is a $4 \times
4$ matrix $P$. The quantum mechanical time evolution of $P$ is
contained in Eq. \densitymateq\ while, in the context of
generalization of \qm , it is contained in Eq. \densitymateqnew\ .

  Let us first describe the quantum mechanical time dependence of $P$.
We find
$$ \eqalign{
P^\qmg (&\tau_1,\tau_2) ={1 + 2\Re(\ek_S \ek_L)\over 2}
\Bigl(\rs^\qmg\otimes\rl^\qmg e^{-\gs\tau_1} e^{-\gl\tau_2} +
\rl^\qmg\otimes \rs^\qmg
         e^{-\gl\tau_1}e^{-\gs\tau_2}  \cr  &\hskip 0.05in
-\ri^\qmg\otimes\rib^\qmg e^{-i\dm (\tau_1-\tau_2)}e^{-\gb(\tau_1+\tau_2)}
 -\rib^\qmg\otimes\ri^\qmg
 e^{+i\dm (\tau_1-\tau_2)}e^{-\gb(\tau_1+\tau_2)}\Bigr) \ .\cr}
\eqn\phidensitytime$$

As in the one kaon system, any observable is obtained by tracing the
density matrix with a suitable hermitian operator.
 The basic observables computed from $P$ are double differential
decay rates, the probabilities that the kaon with momentum $p$
decays into the final state $f_1$ at proper time $\tau_1$ while the
kaon with momentum $(-p)$ decays to the final state $f_2$ at
proper time $\tau_2$.  We denote this quantity as  $\P(f_1,\tau_1;
f_2,\tau_2)$.  If we denote the expression \phidensitytime\ schematically
as $   P = \sum  A_{ij}\, \rho_i \otimes \rho_j $
where $i,j$ run over $S,L,I, \bar I$, and write the corresponding
eigenvalues as  $\lambda_i$,
then the double decay rate is given by
$$ \P(f_1,\tau_1;f_2,\tau_2) =
    \sum_{i,j}  A_{ij} \, \tr[\rho_i \O_{f_1}] \tr[\rho_j \O_{f_2}]
             e^{-\lambda_i \tau_1 - \lambda_j \tau_2} \ .
\eqn\Peval
$$

A situation of particular importance is
the decay into two identical final states $f_1=f_2$, the double decay
rate takes the simple form\refmark\pecceiall
$$  \P^\qmg(f,\tau_1;f,\tau_2) =  C \times \bigl[
      e^{-\gs\tau_1-\gl\tau_2} + e^{-\gl\tau_1 - \gs\tau_2}
  - 2 \cos(\dm (\tau_1-\tau_2)) e^{-\gb(\tau_1+\tau_2)}\bigr]\ .
\eqn\doubleqm
$$
 This quantity has no dependence on the \cp\ and \cpt\ parameters and
depends on the two times in a manner completely
fixed by quantum mechanics irrespective of the properties of the
decay amplitudes.  In particular, at $\tau_1 = \tau_2$, the
double distribution {\it vanishes}, as a consequence of the antisymmetry
of the initial state wavefunction which is preserved in the evolution
of the beam.

The previous steps can be performed taking into account violation of
\qm, and expressions corresponding to \phidensitytime\ and \doubleqm\
are provided in ref. \us. In addition to the natural replacement
$\rk^\qmg \rightarrow \rk$, new structures appear in the time
dependence of the system . We will content ourselves to illustrate
these structures in the case both kaons decay to $\pi^-\ell^+\nu$ or
to $\pi^+\ell^-\bar\nu$. These are examples of decays to identical
final states whose quantum mechanical time dependence has the general
structure depicted in \doubleqm . We find, to first order in small
parameters
 $$ \eqalign{
&\P(\ell^\pm,\tau_1;\ell^\pm,\tau_2) =
    {|a|^4\over 8} \cr
\times&\bigg\{(1 \pm 4\Re\, \ek_M)\bigl[e^{-\gs \tau_1 - \gl \tau_2} +
e^{-\gl \tau_1 - \gs \tau_2} - 2 \cos\dm(\tau_1-\tau_2) e^{-(\gb +
\alpha- \gamma)(\tau_1 + \tau_2)}\bigr]\cr
&\pm 4{\beta\over |d|}\sin(\dm \tau_1 - \phi_{SW})e^{-(\gb +
\alpha-\gamma)\tau_1 }e^{-\gs\tau_2} + (1 \leftrightarrow 2)\cr
&\pm 4{\beta\over |d|}\sin(\dm \tau_1 + \phi_{SW})e^{-(\gb +
\alpha-\gamma)\tau_1 }e^{-\gl\tau_2} + (1 \leftrightarrow 2)\cr
 &+2 {\alpha\over \dm} \sin\dm(\tau_1+\tau_2) \, e^{-(\gb + \alpha -
\gamma)(\tau_1 + \tau_2)}
+ 2 {\gamma\over \dg}\bigl[ e^{-\gl(\tau_1 + \tau_2)} -
e^{-\gs(\tau_1+\tau_2)}\bigr]\bigg\}\cr}
\eqn\bigexpll
 $$
The first term in the brackets has a form quite close to the canonical
form \doubleqm\ predicted by quantum mechanics, while the remaining
terms give systematic corrections to this result. The three following
lines contain totally new dependence.  These new terms signal the
breakdown of the antisymmetry of the final state wave function, that
is, the breakdown of angular momentum conservation. This is expected
in the framework of density matrix evolution equations, as was
explained in ref. \bps. However, in the \phif\ experiments, one does
not need to wait for the problems of energy-momentum conservation to
built up to a macroscopic violation; one can instead track these
violations directly in the frequency dependence of corrections.

The above peculiar dependence on $\tau_1$ and $\tau_2$ is a unique
signature of violation of \qm\ and provides an unambiguous method to
isolate the EHNS parameters from the \cpt\ violating perturbations
of the decay rates from within \qm\ introduced in \Operatordecay\ --
\Ydefs .

One can, for instance, interpolate the double decay rates into
identical final states $\P(f,\tau_1;f,\tau_2)$ on the line of equal
time $\tau_1=\tau_2$. This quantity vanishes identically according to
the principles of \qm\ and thus is of order $\alpha$, $\beta$ and
$\gamma$. As an illustration, the semileptonic double decay rate
at equal time yields
$$  \eqalign{
\P(\ell^\pm,\tau;\ell^\pm,\tau)&/\P(\ell^\pm,\tau;\ell^\mp,\tau) = \cr
& {1\over2}\bigl[1- e^{-2(\alpha-\gamma)\tau}\bigl(1-{\alpha\over \dm}\sin
2\dm\tau\bigr)\bigr]\cr
+  & {1\over2}{\gamma\over \dg}\bigl[ e^{+\dg\tau} - e^{- \dg\tau}\bigr]\cr
\pm  & 4 {\beta\over |d|}\bigl[\sin( \dm\tau-\phi_{SW})e^{-\dg\tau/2} +
\sin( \dm\tau+\phi_{SW})e^{+\dg\tau/2}\bigr]\cr}
\eqn\Qratll$$

The three coefficients $\alpha$, $\beta$, and $\gamma$ are selected by
terms which are monotonic in $\tau$, oscillatory with frequency
$\dm$,and oscillatory with frequency $2\dm$.

There seems to be no difficulty in constraining \cpt\ violation from
outside \qm\ in a \phif\ independently of other \cpt\ violating
perturbation of \qm\ . However, the reverse is not true. Any
observable at a \phif\ is expected to receive $\alpha$, $\beta$
and $\gamma$ corrections. These corrections are, however, easily
computed and can be systematically taken into account using the
knowledge gained on these parameters using methods of the type
described above.

As an illustration, we present the corrections to the formula
predicting the value of $3\Re \ek'/\ek$ using the integrated time
distribution\refmark\pecceiall at fixed $\Dt =\tau_1-\tau_2$ of the
asymmetric decay into charged and neutral pions
 $$   \PP(\Dt) = \int^\infty_{|\Dt|} d(\tau_1 + \tau_2) \,
 \P(\pi^+\pi^-;\pi^0\pi^0;\Dt)\ .
\eqn\Pdiff
 $$
This time interval distribution is very useful for obtaining the
standard \cpv\ parameters of the neutral kaon system. In particular,
one predicts\refmark\dunietz
$$\lim_{|\Dt|\gs \gg 1}{\PP^\qmg(\Dt>0)- \PP^\qmg(\Dt<0) \over
\PP^\qmg(\Dt>0)+ \PP^\qmg(\Dt<0) } = 3 \Re \,
{\ek'\over\ek}.
\eqn\qmeprime
$$
We find instead\foot{ A more exact formula, as well as a more
complete discussion, is given in ref. \us .}
 $$\eqalign{
\lim_{|\Dt|\gs \gg 1}{\PP(\Dt>0)- \PP(\Dt<0) \over
\PP(\Dt>0)+ \PP(\Dt<0) } \simeq
&3\Re \, \ek'/\ek\,\,\biggl[
1- {\gamma \over |d| |\etapp|^2} + 2{\beta\over
|d||\etapp|}     \biggr] \cr
-&3\,\Im \, \ek'/\ek\,\,\biggl[ 2{\beta\over |d||\etapp|}\biggr] \cr }\,
{}.
\eqn\eprimeapprox
 $$
 It is only under the assumption $\Im \ek'/\ek \ll \Re \, \ek'/\ek
\,\times\,\, \bigl(|d||\etapp|/\beta\bigr) $ that \eprimeapprox\ is a
measurement of $ 3 \Re \, \ek'/\ek$.

{\bf Acknowledgements.}

I thank my collaborator M.E. Peskin.
\singlespace
\refout
\vfill\eject
\undertext{FIGURE CAPTION}

{\bf -Fig.1-} Theoretical predictions (a) and experimental data (b).
\vfil \eject
\bye